\documentclass[conference]{IEEEtran}
\IEEEoverridecommandlockouts
\usepackage{cite}
\usepackage{amsmath,amssymb,amsfonts}
\usepackage{algorithmic}
\usepackage{graphicx}
\usepackage{textcomp}
\usepackage{xcolor}
\def\BibTeX{{\rm B\kern-.05em{\sc i\kern-.025em b}\kern-.08em
    T\kern-.1667em\lower.7ex\hbox{E}\kern-.125emX}}

\usepackage{xspace}
\newcommand{\eg}{e.g.\xspace} % Prints ``e.g.'' with proper spacing
\graphicspath{{figures/}{pictures/}{images/}{./}} %

\newcommand{\AppName}[0]{\emph{QVis}\xspace}

\begin{document}

\title{Visual Analytics of Performance of Quantum Computing Systems and Circuit Optimization\\
\thanks{The manuscript is authored by UT-Battelle, LLC under Contract No.~DE-AC05-00OR22725 with the U.S. Department of Energy. The U.S.~Government retains for itself, and others acting on its behalf, a paid-up nonexclusive, irrevocable worldwide license in said article to reproduce, prepare derivative works, distribute copies to the public, and perform publicly and display publicly, by or on behalf of the Government. The Department of Energy will provide public access to these results of federally sponsored research in accordance with the DOE Public Access Plan. 
http://energy.gov/downloads/doe-public-access-plan.}
}

% \author{
%     \IEEEauthorblockN{Junghoon Chae,
%         Chad A.~Steed,
%         Travis S.~Humble
%     }
%     \IEEEauthorblockA{
%         Oak Ridge National Laboratory, Oak Ridge, USA, 
%         \{chaej, steedca, humblets\}@ornl.gov
%     }
% }

\author{\IEEEauthorblockN{Junghoon Chae}
\IEEEauthorblockA{Oak Ridge National Laboratory \\
Oak Ridge, USA \\
chaej@ornl.gov}
\and
\IEEEauthorblockN{Chad A.~Steed}
\IEEEauthorblockA{Oak Ridge National Laboratory \\
Oak Ridge, USA \\
steedca@ornl.gov}
\and
\IEEEauthorblockN{Travis S.~Humble}
\IEEEauthorblockA{Oak Ridge National Laboratory \\
Oak Ridge, USA \\
humblets@ornl.gov}
}

\maketitle

\begin{abstract}
Driven by potential exponential speedups in business, security, and scientific scenarios, interest in quantum computing is surging. This interest feeds the development of quantum computing hardware, but several challenges arise in optimizing application performance for hardware metrics (e.g., qubit coherence and gate fidelity). In this work, we describe a visual analytics approach for analyzing the performance properties of quantum devices and quantum circuit optimization. Our approach allows users to explore spatial and temporal patterns in quantum device performance data and it computes similarities and variances in key performance metrics. Detailed analysis of the error properties characterizing individual qubits is also supported. We also describe a method for visualizing the optimization of quantum circuits. The resulting visualization tool allows researchers to design more efficient quantum algorithms and applications by increasing the interpretability of quantum computations.
\end{abstract}


\section{Introduction}
\begin{figure}[b]
\centering
\includegraphics[width=.84\columnwidth]{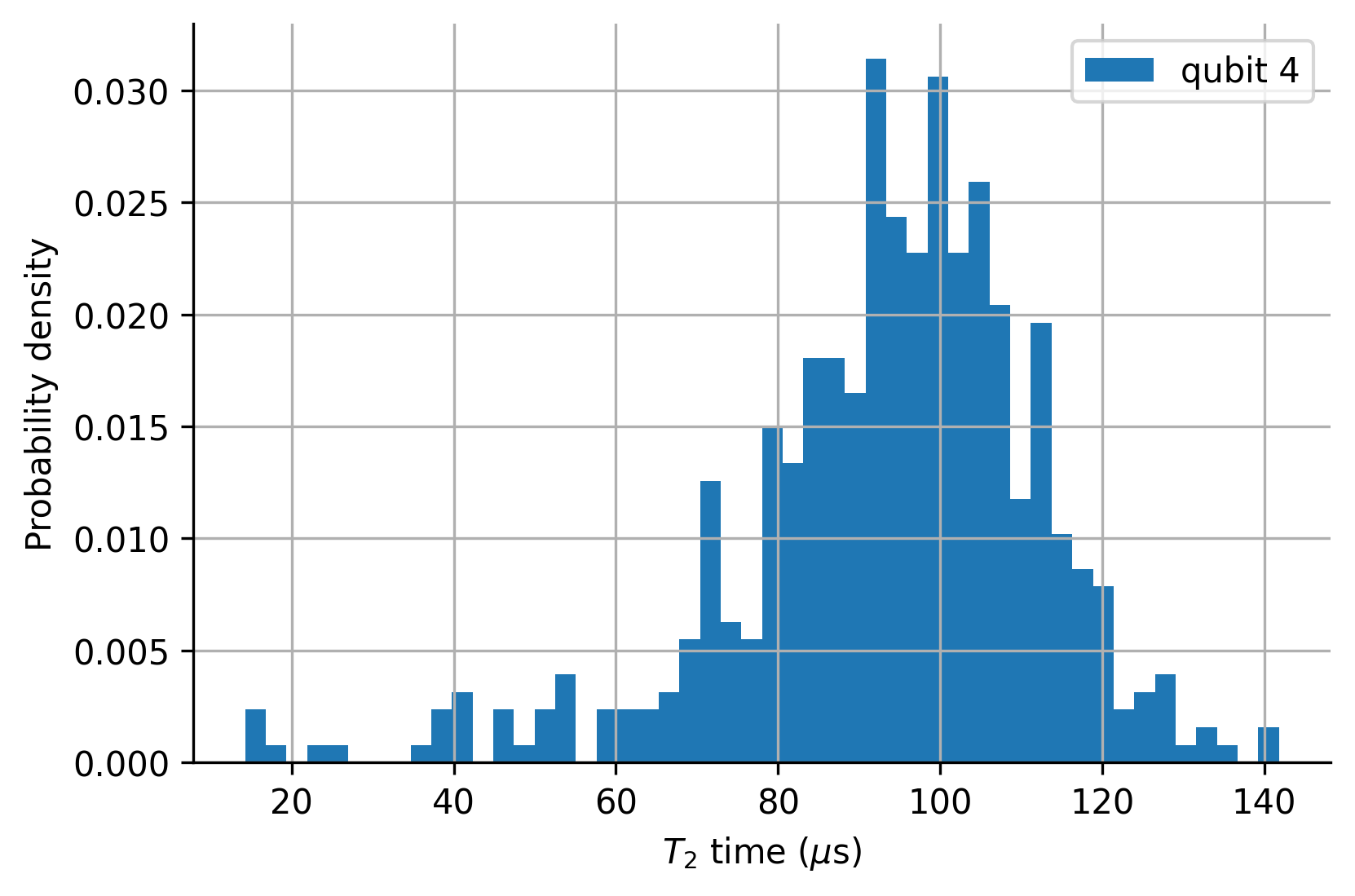}
\caption{The distribution of $T_2$ times observed for qubit 4 of the IBM transmon device Washington for the period 1-Jan-2022 to 30-Apr-2023. The distribution of ${T_2}$ underlies variations in system behavior and fluctuations in computational errors that can be revealed through visual analytics.
}
\label{fig:T2_histogram}
\end{figure}

Quantum computing devices have the potential to revolutionize computation in many critical domains. However, current devices suffer from significant noise, including decoherence ($T_2$), that results in high error rates during computations. For example, Figure~\ref{fig:T2_histogram} illustrates the probability distribution of the $T_2$ time for a qubit of the transmon device named Washington based on daily data from IBM's published characterizations from 1-Jan-2022 to 30-Apr-2023. The $T_2$ metric quantifies the duration before a quantum superposition state transitions into a classical state. In the quest to improve quantum devices and reduce such errors, understanding the sources of noise is essential.

As the complexity of quantum device architectures increases, reasoning about noise and its impact on device performance becomes more difficult~\cite{lilly2020modeling,lotshaw2022scaling,dasgupta2020characterizing,dasgupta2022assessing}. To uncover insights about noisy quantum device behaviors, more advanced analysis tools are required. Optimizing quantum circuits to meet the specifications of a particular quantum device extracts the most value from the given hardware and increases both the accuracy of the results and the scalability of the circuits. Also, creating shallower (low-depth) circuits by circuit optimization could reduce the execution cost of cloud quantum computers. However, many techniques for optimizing quantum circuits address different aspects of circuit efficiency and performance, such as reordering, fusion, and elimination of gates and qubit mapping. New optimization strategies may emerge as quantum devices evolve. Developers need new tools to interpret and understand the optimization results as they strive to design more efficient algorithms.

%%%%%%%%%%%%%%%%%%%%%%%%%%%%%%%%%
%%%%%%%%%%%%%%%%%%%%%%%%%%%

%%%%%%%%%%%%%%%%%%%%%%%%%%%%%%%%%
\begin{figure*}[t]
\centering
\includegraphics[width=1.0\textwidth]{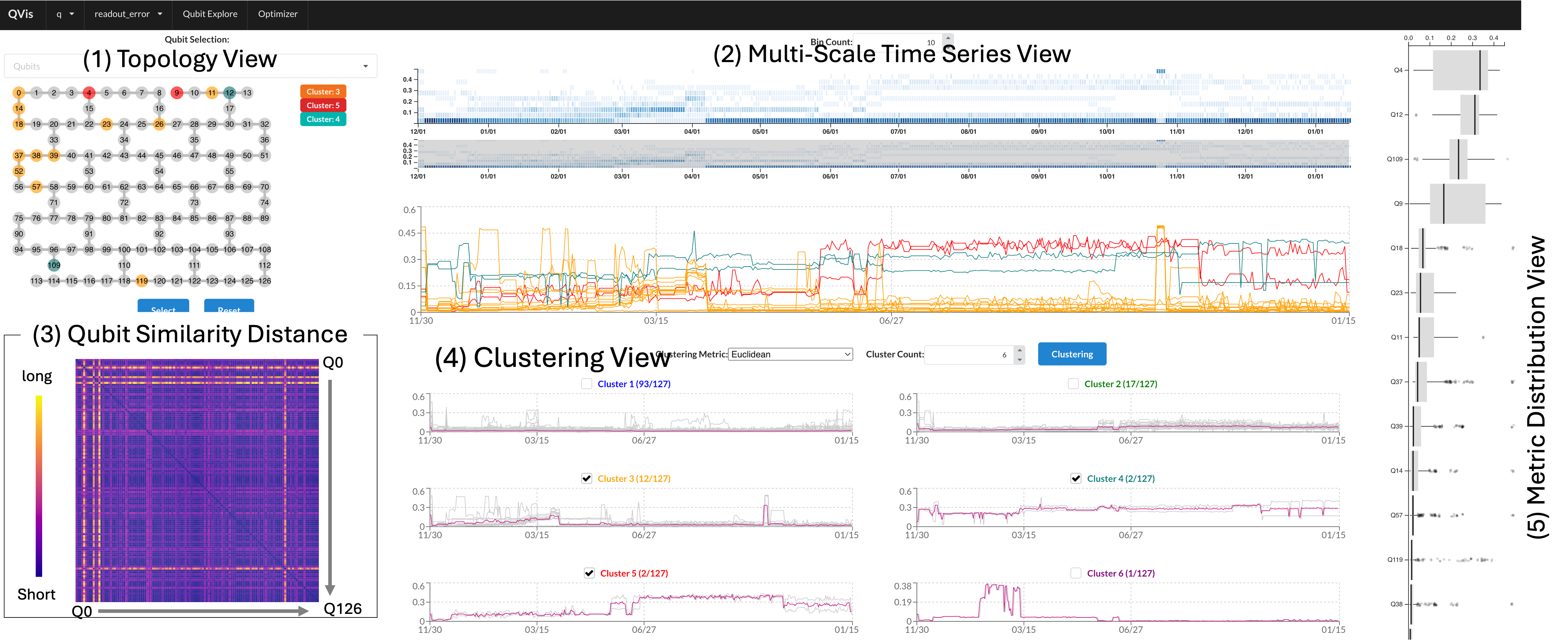}
\caption{\AppName consists of several visualizations including Topology View (1), Multi-Scale Time Series View (2), Qubit Similarity Distance (3), Clustering View (4), and Metric Distribution View (5). Each view supports different analytics but these visualizations are tightly interconnected to support holistic analysis.}
\label{fig:system_overview}
% \vspace{-0.5cm}
\end{figure*}
%%%%%%%%%%%%%%%%%%%%%%%%%%%%%%%%%

In this paper, we address the challenge of analyzing noise and error of quantum devices and examining circuit optimization results through the development of a visual analytics tool called \AppName. Our tool enables human-directed analysis of performance metrics of a quantum device and visualization of optimization results. It provides a dashboard consisting of several interactive visualization components that support temporal and statistical performance trend analysis, clustering of the performance behavior of quantum bits (qubits), and a comprehensive understanding of key performance metrics and topology graphs of the qubits. \AppName also allows users to run circuit optimizations and visually explore the results. 
% These capabilities help developers gain new insights and design more efficient algorithms and better utilization of quantum resources.
Through these visualizations, developers can analyze and interpret optimized circuits to gain insights into quantum algorithms and design more efficient algorithm. This leads to better utilization of quantum resources and ultimately improved performance of quantum computations. We demonstrate the applications of these features based on the analysis of a 127-qubit data set derived from the IBM Washington processor over 16 months.

% \section{Background}
% Background Knowledge of Data

\section{Related Work}

Visualization of quantum computing is an emerging area of research that aims to bridge the gap between complex quantum computing concepts and the practical needs of quantum algorithm developers~\cite{bethel2023quantum}. Visualizing quantum computing processes and results also plays a critical role in understanding and developing quantum algorithms. Several research efforts have focused on developing effective visualization techniques for quantum computing. This section reviews significant contributions to the field, highlighting various tools and methodologies developed to enhance understanding and accessibility.

\subsection{Visualization for Quantum Circuits}
Quantum circuit diagrams are a standard way to represent quantum algorithms and computations. Interactive visualizations of quantum circuits can aid in understanding their structure and behavior. IBM's Quantum Composer~\cite{QuantumComposer} provides live visualizations of quantum circuits, including state vectors, probabilities, and q-spheres, allowing users to explore the effects of gates and operations on qubit states. Cirq~\cite{Cirq} is another prominent tool in this domain, providing a Python library for designing, simulating, and optimizing quantum circuits. Cirq includes features for visualizing quantum circuits, which help users understand the flow and transformation of qubit states through various gates. Quantivine~\cite{wen2024quantivine} presents a novel approach to visualizing quantum circuits by utilizing semantic analysis to enhance comprehension and readability. Traditional quantum circuit diagrams face challenges with scalability and readability, especially as circuits grow in complexity. Quantivine addresses these issues by integrating semantic structures, called abstract syntax trees (AST) and meanings into the visual representation of quantum circuits, making them more intuitive and easier to analyze. 

\subsection{Quantum State and Entanglement Visualization}
Understanding the evolution of quantum states and entanglement is essential for quantum algorithm development. The Bloch Sphere is a widely used tool for visualizing quantum states~\cite{altepeter2009multiple}. However, it falls short in visualizing quantum entanglement and superposition. To overcome this limitation, VENUS~\cite{ruan2023venus} was proposed for quantum state representation. It is a geometric representation for visualizing quantum states of single and entangled qubits. VENUS uses 2D shapes like semicircles to encode probability distributions and superposition, allowing users to explore quantum entanglement. The tools we have developed focus on showing and analyzing the optimized results of quantum circuits, rather than the ability to write quantum circuits or visualize quantum states. Bley et al.~\cite{PhysRevResearch.6.023077} explore methods for representing and visualizing a new perspective on entanglement in few-qubit systems. The authors utilize Dimensional Circle Notation (DCN), an extension of circle notation, to depict quantum states in an n-dimensional space. This visual representation aids in understanding the symmetry conditions that correspond to separable states, thereby offering insights into entanglement properties that are often not intuitive through traditional mathematical descriptions.

%%%%%%%%%%%%%%%%%%%%%%%%%%%%%%%%%
\begin{figure*}[t]
\centering
\includegraphics[width=1.0\textwidth]{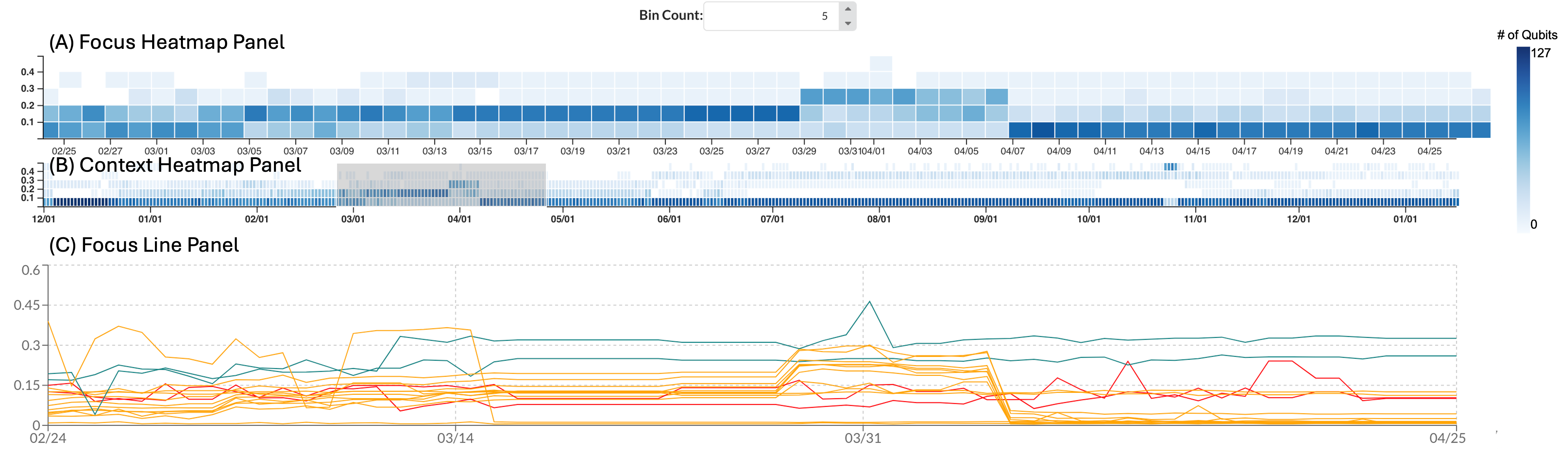}
\caption{Multi-scale Time Series View features three linked temporal visualizations: Focus Heatmap (A), Context Heatmap (B), and Focus Line Panels. Panels (A) and (B) work together as a focus+context visualization using a binned representation of the selected device metric. The focus line panel displays the focus time range metric value as a line chart. Currently, the readout error metric is displayed in this view.}
\label{fig:heatmap_readouterror}
\vspace{-0.5cm}
\end{figure*}
%%%%%%%%%%%%%%%%%%%%%%%%%%%%%%%%%

\subsection{Visualization of Quantum Device Performance}
Most existing studies related to quantum noise have focused on mitigating errors and noise of quantum devices~\cite{cai2023quantum}. Harper et al.~\cite{harper2020efficient} introduced an efficient protocol for learning quantum noise and constructed a quantum noise correlation matrix, allowing easy visualization of noise correlations between all qubit pairs. This enabled the discovery of long-range two-qubit noise correlations that were previously undetected. VACSEN~\cite{ruan2022vacsen} is a visualization approach proposed to achieve noise-aware quantum computing. It provides a holistic picture of quantum noise through multiple coordinated views to overview temporal noise evolution across quantum computers. 

In this work, we focus on helping users analyze qubits in a device from different perspectives, taking into account errors and noise and other performance metrics and topological information, and applying data science techniques to classify them. Our tool does not show only one aspect of quantum errors but rather provides a holistic view that allows exploring multiple aspects of the performance values of qubits and categorizing them according to their performance behavior. These results extend prior efforts \cite{steed2023qvis} through  finer granularity and control over the visualization of performance data.  

\section{Quantum System Performance Data}
The dataset analyzed in this paper was derived from a subset of the Washington device performance characterization data collected during a $16$-month period starting on 1-Jan-2022 and ending on 30-Apr-2023~\cite{onlinedataset}. The data set includes performance metrics including state preparation and measurement (SPAM) error rate, gate error rate, gate duration, qubit lifetime $(T_1)$ and qubit coherence time $(T_2)$ time. For demonstration of our analysis using \AppName in this paper, we use the read-out error and daily $T_1$ and $T_2$ data. $T_1$ and $T_2$ metrics quantify the performance of the register to store quantum information with higher values being more favorable.

\section{\AppName Dashboard Overview}
As shown in Figure~\ref{fig:system_overview}, users can select a specific performance metric of the data using the drop-down menus at the top. Also, there are two tab menus at the top of the tool: Qubit Explore and Optimizer. The Qubit Explore tab is for qubit performance metric exploration. It consists of the following multiple coordinated visualization components: Topology view (1), Multi-Scale Time Series view (2), Qubit Similarity Distance view (3), Clustering view (4), and Metric Distribution view (5). Each view visualizes different aspects of qubit performance data. In this case, the views present the read-out error data of the quantum device. The interconnected views provide highly interactive capabilities for effective analysis. The optimizer tab in Figure~\ref{fig:optimization} also has multiple visualization components for visualizing quantum circuit optimization results and metrics.

%%%%%%%%%%%%%%%%%%%%%%%%%%%%%%%%%
\begin{figure}[b]
\centering
\includegraphics[width=1.0\columnwidth]{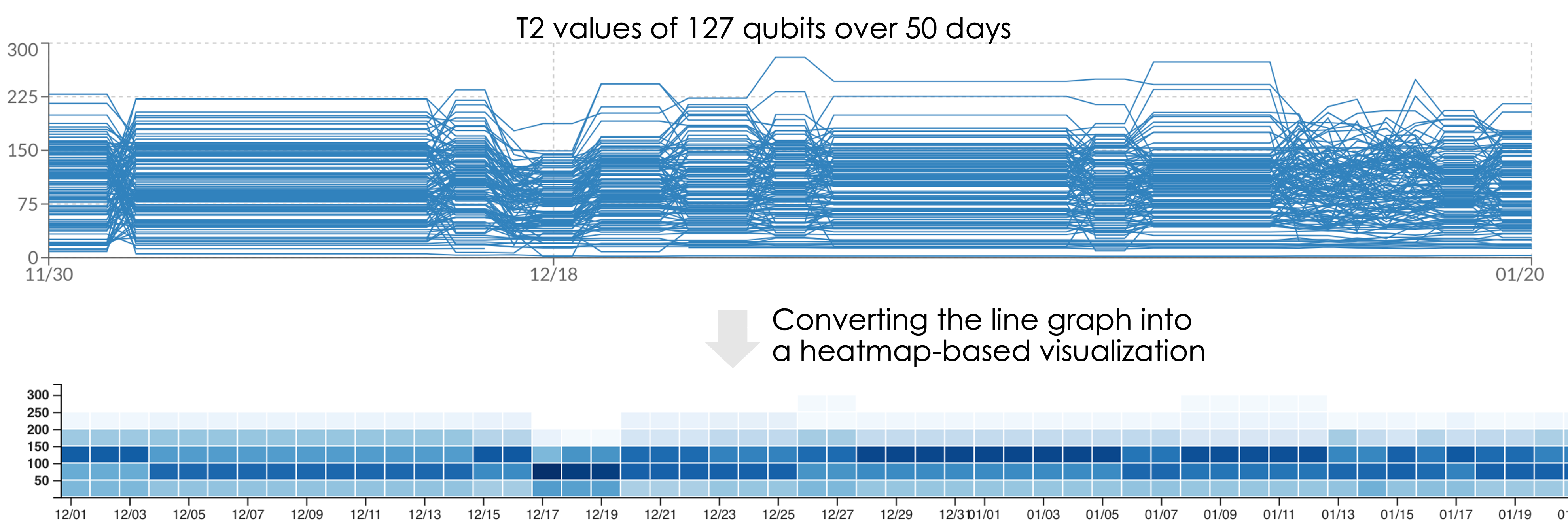}
\caption{Aggregated representation of temporal data points to mitigate a visual over-plotting issue.}
\label{fig:heatmap}
% \vspace{-0.5cm}
\end{figure}
%%%%%%%%%%%%%%%%%%%%%%%%%%%%%%%%%

%%%%%%%%%%%%%%%%%%%%%%%%%%%%%%%%%
\begin{figure*}[t]
\centering
\includegraphics[width=1.0\textwidth]{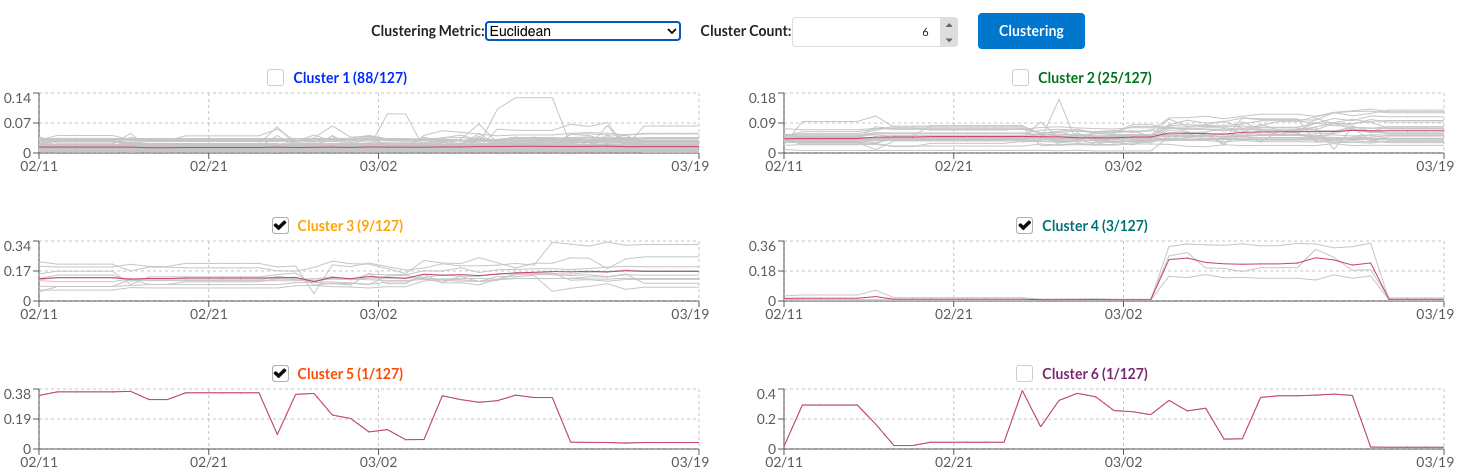}
\caption{Clustering View: Each line graph represents one clustering. Users can choose the distance metric and how many clusters they want. The different colors of each cluster title make it possible to distinguish between clusters in other views.}
\label{fig:clustering}
% \vspace{-0.5cm}
\end{figure*}
%%%%%%%%%%%%%%%%%%%%%%%%%%%%%%%%%

\section{Visual Analytics of Qubit Performance}
\subsection{Multi-scale Temporal Performance Exploration}
The Multi-scale Time Series View provides three interactive visualization panels for exploring temporal variability in quantum device data in Figure~\ref{fig:heatmap_readouterror}. The read-out errors are displayed in each panel, with the collection time mapped to the $x$-axis and the metric value to the $y$-axis. It is important to note that although we show only the read-out error metric in the figure, the user can choose to view any available metrics. 

The focus+context visualization approach is often used when analyzing temporal information because it allows users to see a smaller time range (the focus) and a wider time range (the context) simultaneously. The focus+context visualization panels in Figure~\ref{fig:heatmap_readouterror} show the data for the selected time range of interest in the focus panel while also showing the focus time range within the context of the overall data set in the lower context panel. 

A time range of interest can be magnified by dragging a selection in the context panel, shown as the gray rectangular region in Figure~\ref{fig:heatmap_readouterror} (B). This action forces the focus view to redraw using the selected time range. The user can also fine-tune the focus range selection by dragging either the start or end limit or by panning the entire time range selection. Here, the daily granularity of the data sets the minimum range. These interactions allow users to flexibly navigate to and investigate various time ranges of interest at different scales while preserving contextual awareness of the whole.

The focus+context panels display heatmaps, visual representations of the aggregated data to avoid over-plotting and clutter issues that are often associated with displaying many time series records in a single plot. The heatmap is constructed by partitioning the visible portion of the data into two-dimensional bins in both the $y$- and $x$-axis dimensions. In Figure~\ref{fig:heatmap}, for example, the line graph (top) shows $T2$ values of the 127 qubits. While visually cluttered and overlapping makes it difficult to see patterns accurately, the heatmap (bottom) can help alleviate visual overflow. We partitioned the data into two-dimensional bins in the y-value and x-time dimensions. The darkness of the color of each bin represents the number of qubits that correspond to the bin. It makes it easier to see the detailed changes in the values.

In Figure~\ref{fig:heatmap_readouterror}, the presentation of the readout error data yields insights into how coherence changes over the selected time range. The user can use menu components to adjust the bin count to refine the analysis. After the bin count is set, daily time series records for each qubit are processed and assigned to the appropriate bins based on the time and metric values. When the binning process is complete, each bin contains associated qubits. The bins are visualized in the focus+context panels using the color scale shown on the right side of the panels. The color scale indicates the number of qubits associated with each bin, with darker blues indicating a larger number of qubits. Although the binned representation sacrifices the display of individual qubit values to maximize legibility, the contextual display preserves connections to the full data set and reveals broader temporal patterns. Also, users can hover over a bin to access statistical summaries (\eg, the number of qubits, median value) in the form of a textual tooltip. 
% A menu interface (not shown in Figure~\ref{fig:heatmap_readouterror}) allows the user to specify the metric and specific qubits to generate customized views.

\subsection{Clustering Analysis of Qubits}
In addition to multi-scale temporal analysis, the vision for \AppName includes automated analytical methods that guide the user to potentially significant insights. \AppName's initial analytical offering focuses on the clustering of qubits in terms of their temporal performance. By clustering key quantum hardware performance metrics in time of qubits, \AppName supports understanding the behaviors of the device, extracting significant patterns, and identifying outliers. We utilize a k-means clustering algorithm for time series data~\cite{JMLR:v21:20-091}. 

Our system supports multiple types of distance metrics, such as Euclidean and Dynamic Time Warping~\cite{petitjean2011global}. We compute the distance between the \textit{i}th point of one series and \textit{i}th point of another using the selected metric. Despite some limitations, such as being invariant to time shift, the approach using Euclidean distance supports grouping the time series data into daily behaviors (\eg, average value of each date). Additionally, the time series records for each qubit are of equal lengths, meaning we can avoid using Dynamic Time Warping (DTW)~\cite{petitjean2011global}, which is a similarity measure for variable length time series that would result in significant additional computation. The k-means clustering algorithm constructs clusters of data by splitting samples into $k$ groups and minimizing the sum of squares in each cluster. Through trial and error, we decided to use $6$ for the \textit{k} parameter by default, but this parameter can be changed depending on the user's analytics objectives. 

The Clustering View in Figure~\ref{fig:clustering} shows the clustering results for the read-out error metric data. Each subplot represents a cluster, where the $x$-axis is mapped to time and the $y$-axis is mapped to the error value. The semi-transparent grey lines show the individual instances in the cluster, and the red lines show each cluster's barycenter. The barycenter is the arithmetic mean for each point in time where the summed Euclidean distance is minimized for each of them. This view not only shows the clustering results but also allows the user to select clusters of interest, and the different aspects of the qubits in that cluster are shown in other views.

%%%%%%%%%%%%%%%%%%%%%%%%%%%%%%%%%
\begin{figure}[t]
\centering
\includegraphics[width=0.8\columnwidth]{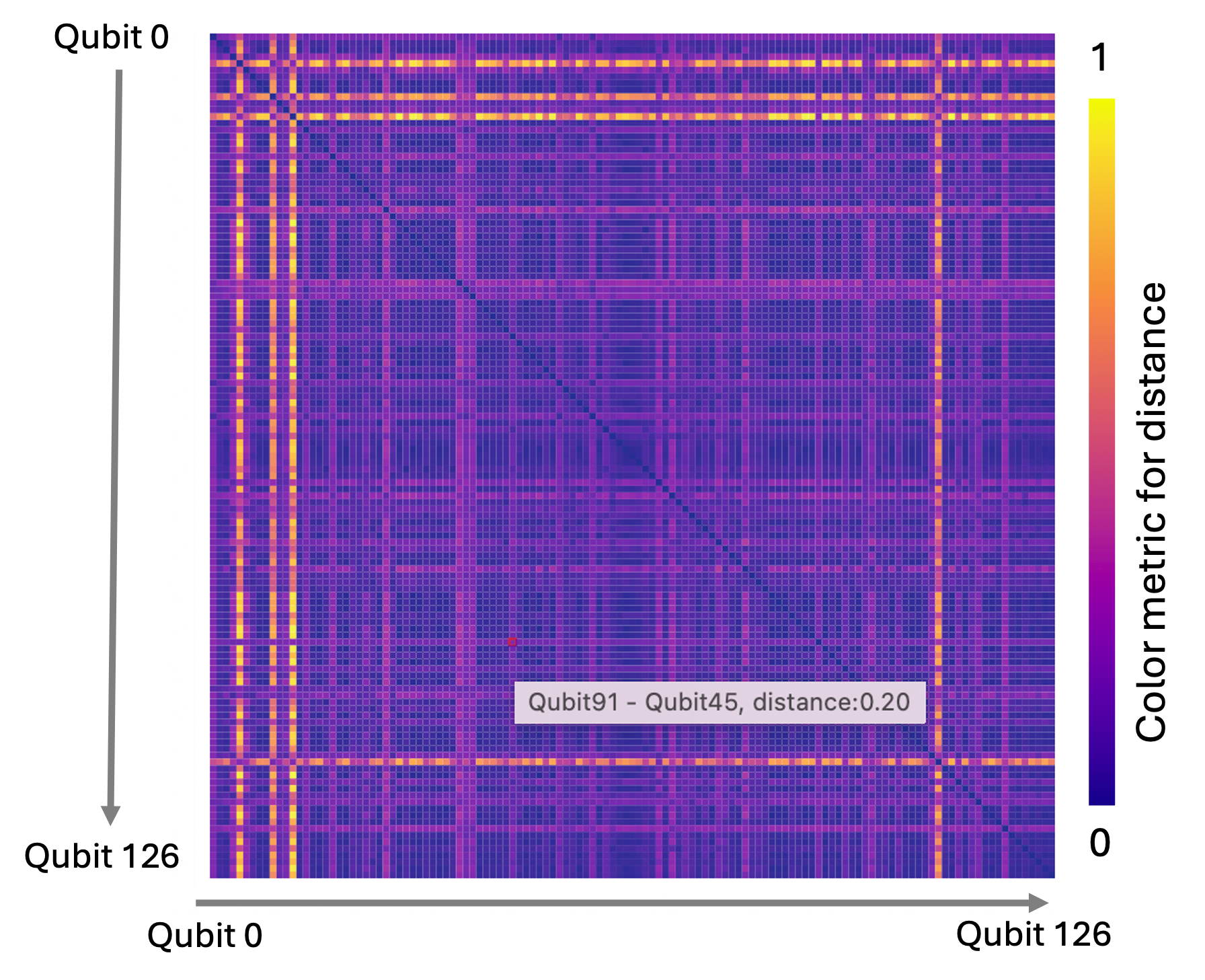}
\caption{Qubit Similarity Distance View: Each cell represents the similarity distance of readout error between two qubits.
}
\label{fig:matrix}
\vspace{-0.5cm}
\end{figure}
%%%%%%%%%%%%%%%%%%%%%%%%%%%%%%%%%

\subsection{Similarity Distance View}
The similarity distances of all qubit pairs in terms of the selected hardware data are computed using the selected distance metric during the k-mean clustering process. As shown in Figure~\ref{fig:matrix}, the distance matrix is visualized using a heatmap representation. The color of each cell represents the distance. Bright yellow is mapped to the maximum distance and dark blue to the minimum. The heatmap facilitates comparative analysis. For the distance matrix, similar items will have similar colors, making it easy to compare the distances between multiple pairs of items simultaneously. Also, the heatmap provides an interpretable visual representation of the data. The color gradients reveal patterns, clusters, and outliers at a glance. For example, for this result, we can see that the four qubits (4, 9, 12, and 109) have a distinctly different pattern than the other qubits. Clustering can be used to classify qubits, but showing the similarity of all pairs at a glance can efficiently filter out the qubits with abnormal patterns.

\subsection{Interactive Hardware Topology Visualization}
A comprehensive understanding of the performance metrics and topology graphs of the qubits is an important step in circuit design. The Topology View shows the layout and connectivity of the qubits (circles) of the quantum computing machine in Figure~\ref{fig:system_overview} (top-left). Rather than simply showing the topology of qubits, this view works with other views to provide interactive analytical capabilities. When you select a specific cluster in the clusters view, the color of the qubits corresponding to the selected cluster changes to match the color of the cluster title. At the same time, other views are updated to show qubit information for the selected cluster. 

On the other hand, in the topology view, users can select specific qubits. Only data of the selected qubits are displayed in the other views. For example, as shown in Figure~\ref{fig:system_overview}, clusters 3, 4, and 5 are selected and the qubits of the selected qubits are highlighted by the corresponding colors in the topology view (1). As we mentioned in the previous section, the qubits (4, 9, 12, and 109) with noticeably different patterns from other ones belong to clusters 4 and 5.

\section{Visualizing Quantum Circuit Optimization}
After designing a circuit, developers use a tool to map a circuit to a specific quantum device based on the specs of the quantum hardware. This process is called transpilation and optimizing the circuit helps extract the best performance from the quantum system. Optimization can reduce the quantum resources needed, such as the number of qubits and gates, and reduce the run time costs. It can also increase the accuracy of the results because the shallower the depth of the circuit, the less chance of noise and errors due to qubits and gates. Optimization  greatly affects the performance of the circuit running on the hardware.

\begin{figure*}[t]
\centering
\includegraphics[width=1.0\textwidth]{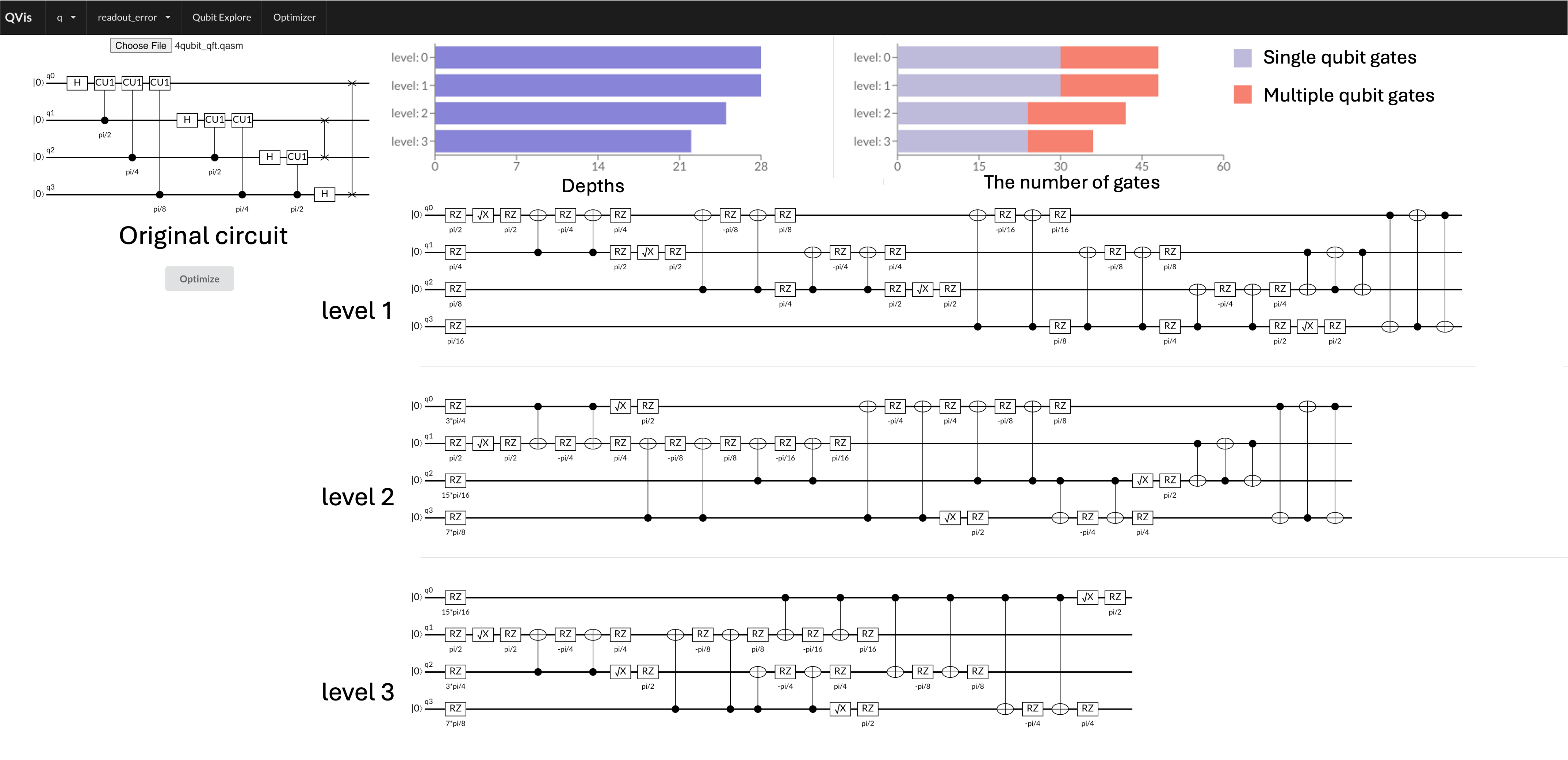}
\caption{Visual Analytics of Quantum Circuit Optimization: Users can load a quantum circuit in QASM format. Depths and the number of gates (local and non-local) of the optimized circuits are displayed as bar charts respectively. It also displays the optimized circuits themselves.}
\label{fig:optimization}
\vspace{-0.5cm}
\end{figure*}

We have integrated \AppName with the IBM Qiskit transpiler~\cite{Qiskit} to create a method for visualizing the effects of different optimizations. There are three optimization levels, ranging from 1 to 3, where a higher value indicates more time in seeking an optimal implementation of the circuit, e.g., using fewer gates. As shown in Fig.~\ref{fig:optimization}, users can load an input circuit (top-left) in Quantum Assembly Language (QASM) format, \AppName calls the transpilation and optimization processes and displays the optimization metrics: circuit depth and the number of gates. \AppName shows the circuit depths for each level as a histogram. It also shows the number of single and multiple qubit gates as a stacked histogram. Qubit gates can be categorized as single and multiple qubit gates. Single qubit gates act on individual qubits and are represented by unitary matrices. Multiple qubit gates act on two or more qubits simultaneously, allowing for entanglement and interactions between qubits. Reducing multi-qubit gates typically has a more significant impact on the overall circuit performance because multi-qubit operations are more error-prone and resource-intensive.

Finally, \AppName displays the optimized deconstructed version of circuits for each optimization level. Through these visualizations, developers can analyze optimized circuits to design more efficient quantum algorithms, ultimately improving the performance of their algorithms.

%%%%%%%%%%%%%%%%%%%%%%%%%%%%%%%%%
%%%%%%%%%%%%%%%%%%%%%%%%%%%%%%%%%

% \section{Discussion}
% Dataset
% Limitations
% - circuit visualization
% - Integrating other quantum computing framework

\section{Conclusions}
\AppName provides new human-centered visualizations to explore how noise and errors manifest in quantum computing devices. In addition to the visualization techniques, in this paper, we have demonstrated the application of temporal clustering to identify multiple subsets of qubits that demonstrate similar temporal patterns. We also described \AppName's integration with an external framework for optimizing quantum circuits to provide visualizations for analyzing and understanding the optimization results. \AppName is a promising tool for informing device performance studies and revealing insights into processor conditions that support reliable behavior.

Future work will extend the temporal and multivariate analysis techniques to incorporate topological analytics. This will  include the development of automated methods to reveal correlations across performance metrics. We envision a tightly coupled visual analytic system that is publicly accessible and dynamically pulls performance metrics for multiple quantum computing devices to enable monitoring of system performance. These tools are an important advance in understanding the stability and reliability of noisy quantum computing devices and optimal circuit designs.

\section*{Acknowledgment}
This research was supported by the US Department of Energy, Advanced Scientific Computing Research (ASCR), Accelerated Research in Quantum Computing program.

\bibliographystyle{IEEETran}

\bibliography{IEEEabrv,qvis}

\begin{thebibliography}{10}
\providecommand{\url}[1]{#1}
\csname url@samestyle\endcsname
\providecommand{\newblock}{\relax}
\providecommand{\bibinfo}[2]{#2}
\providecommand{\BIBentrySTDinterwordspacing}{\spaceskip=0pt\relax}
\providecommand{\BIBentryALTinterwordstretchfactor}{4}
\providecommand{\BIBentryALTinterwordspacing}{\spaceskip=\fontdimen2\font plus
\BIBentryALTinterwordstretchfactor\fontdimen3\font minus
  \fontdimen4\font\relax}
\providecommand{\BIBforeignlanguage}[2]{{%
\expandafter\ifx\csname l@#1\endcsname\relax
\typeout{** WARNING: IEEEtran.bst: No hyphenation pattern has been}%
\typeout{** loaded for the language `#1'. Using the pattern for}%
\typeout{** the default language instead.}%
\else
\language=\csname l@#1\endcsname
\fi
#2}}
\providecommand{\BIBdecl}{\relax}
\BIBdecl

\bibitem{lilly2020modeling}
M.~N. Lilly and T.~S. Humble, ``Modeling noisy quantum circuits using
  experimental characterization,'' \emph{arXiv preprint arXiv:2001.08653},
  2020.

\bibitem{lotshaw2022scaling}
P.~C. Lotshaw, T.~Nguyen, A.~Santana, A.~McCaskey, R.~Herrman, J.~Ostrowski,
  G.~Siopsis, and T.~S. Humble, ``Scaling quantum approximate optimization on
  near-term hardware,'' \emph{Scientific Reports}, vol.~12, no.~1, p. 12388,
  2022.

\bibitem{dasgupta2020characterizing}
S.~Dasgupta and T.~S. Humble, ``Characterizing the stability of nisq devices,''
  in \emph{2020 IEEE International Conference on Quantum Computing and
  Engineering (QCE)}.\hskip 1em plus 0.5em minus 0.4em\relax IEEE, 2020, pp.
  419--429.

\bibitem{dasgupta2022assessing}
------, ``Assessing the stability of noisy quantum computation,'' in
  \emph{Quantum Communications and Quantum Imaging XX}, vol. 12238.\hskip 1em
  plus 0.5em minus 0.4em\relax SPIE, 2022, pp. 44--49.

\bibitem{bethel2023quantum}
E.~Bethel, M.~G. Amankwah, J.~Balewski, R.~V. Beeumen, D.~Camps, D.~Huang, and
  T.~Perciano, ``Quantum computing and visualization: A disruptive
  technological change ahead,'' \emph{IEEE Computer Graphics and Applications},
  vol.~43, no.~06, pp. 101--111, nov 2023.

\bibitem{QuantumComposer}
\BIBentryALTinterwordspacing
IBM. (2024) Quantum composer. [Online]. Available:
  \url{https://https://learning.quantum.ibm.com/tutorial/explore-gates-and-circuits-with-the-quantum-composer}
\BIBentrySTDinterwordspacing

\bibitem{Cirq}
{CirQ Developers}, ``Cirq: A python framework for creating, editing, and
  invoking noisy intermediate scale quantum (nisq) circuits,'' 2023.

\bibitem{wen2024quantivine}
Z.~Wen, Y.~Liu, S.~Tan, J.~Chen, M.~Zhu, D.~Han, J.~Yin, M.~Xu, and W.~Chen,
  ``Quantivine: A visualization approach for large-scale quantum circuit
  representation and analysis,'' \emph{IEEE Transactions on Visualization and
  Computer Graphics}, vol.~30, no.~01, pp. 573--583, jan 2024.

\bibitem{altepeter2009multiple}
J.~B. Altepeter, E.~R. Jeffrey, M.~Medic, and P.~Kumar, ``Multiple-qubit
  quantum state visualization,'' in \emph{2009 Conference on Lasers and
  Electro-Optics and 2009 Conference on Quantum electronics and Laser Science
  Conference}, 2009, pp. 1--2.

\bibitem{ruan2023venus}
S.~Ruan, R.~Yuan, Q.~Guan, Y.~Lin, Y.~Mao, W.~Jiang, Z.~Wang, W.~Xu, and
  Y.~Wang, ``Venus: A geometrical representation for quantum state
  visualization,'' in \emph{Computer Graphics Forum}, vol.~42, no.~3.\hskip 1em
  plus 0.5em minus 0.4em\relax Wiley Online Library, 2023, pp. 247--258.

\bibitem{PhysRevResearch.6.023077}
\BIBentryALTinterwordspacing
J.~Bley, E.~Rexigel, A.~Arias, N.~Longen, L.~Krupp, M.~Kiefer-Emmanouilidis,
  P.~Lukowicz, A.~Donhauser, S.~K\"uchemann, J.~Kuhn, and A.~Widera,
  ``Visualizing entanglement in multiqubit systems,'' \emph{Phys. Rev. Res.},
  vol.~6, p. 023077, Apr 2024. [Online]. Available:
  \url{https://link.aps.org/doi/10.1103/PhysRevResearch.6.023077}
\BIBentrySTDinterwordspacing

\bibitem{cai2023quantum}
Z.~Cai \emph{et~al.}, ``Quantum error mitigation,'' \emph{Reviews of Modern
  Physics}, vol.~95, no.~4, p. 045005, 2023.

\bibitem{harper2020efficient}
R.~Harper, S.~T. Flammia, and J.~J. Wallman, ``Efficient learning of quantum
  noise,'' \emph{Nature Physics}, vol.~16, no.~12, pp. 1184--1188, 2020.

\bibitem{ruan2022vacsen}
S.~Ruan, Y.~Wang, W.~Jiang, Y.~Mao, and Q.~Guan, ``Vacsen: A visualization
  approach for noise awareness in quantum computing,'' \emph{IEEE Transactions
  on Visualization and Computer Graphics}, vol.~29, no.~1, pp. 462--472, 2022.

\bibitem{steed2023qvis}
C.~A. Steed, J.~Chae, S.~Dasgupta, and T.~S. Humble, ``Qvis: A visual analytics
  tool for exploring noise and errors in quantum computing systems,'' in
  \emph{2023 IEEE International Conference on Quantum Computing and Engineering
  (QCE)}, vol.~2.\hskip 1em plus 0.5em minus 0.4em\relax IEEE, 2023, pp.
  211--214.

\bibitem{onlinedataset}
\BIBentryALTinterwordspacing
S.~Dasgupta. (2023) Quantum characterization metrics data set. [Online].
  Available: \url{https://github.com/quantumcomputing-lab/nisqReliability/}
\BIBentrySTDinterwordspacing

\bibitem{JMLR:v21:20-091}
\BIBentryALTinterwordspacing
R.~Tavenard \emph{et~al.}, ``Tslearn, a machine learning toolkit for time
  series data,'' \emph{Journal of Machine Learning Research}, vol.~21, no. 118,
  pp. 1--6, 2020. [Online]. Available:
  \url{http://jmlr.org/papers/v21/20-091.html}
\BIBentrySTDinterwordspacing

\bibitem{petitjean2011global}
F.~Petitjean, A.~Ketterlin, and P.~Gan{\c{c}}arski, ``A global averaging method
  for dynamic time warping, with applications to clustering,'' \emph{Pattern
  recognition}, vol.~44, no.~3, pp. 678--693, 2011.

\bibitem{Qiskit}
{Qiskit contributors}, ``Qiskit: An open-source framework for quantum
  computing,'' 2023.

\end{thebibliography}

\end{document}